\documentclass{aa} 

\bibpunct{(}{)}{;}{a}{}{,} 

\usepackage{graphicx}
\usepackage{txfonts}
\usepackage{bm}
\usepackage{ mathrsfs }
\usepackage{courier}
\usepackage{pgfplots}
\usepackage{pgf}
\usepackage{comment}
\usepackage{natbib}
\usepackage{acronym}
\acrodef{ISM}{interstellar medium}
\acrodef{MHD}{magnetohydrodynamic}
\acrodef{SN}{supernova}
\acrodefplural{SN}[SNe]{supernovae}
\acrodef{GMC}{giant molecular cloud}
\acrodefplural{GMC}[GMCs]{giant molecular clouds}
\acrodef{PDF}{probability distribution function}
\acrodefplural{PDF}[PDFs]{probability distibution functions}
\acrodef{IGM}{intergalactic medium}
\acrodef{ROI}{region of interest}
\acrodefplural{ROI}[ROIs]{regions of interest}
\acrodef{SLUG}{Stochastically Lighting Up Galaxies}

\begin{document}

   \title{Comparing simulated $^{26}$Al maps to gamma-ray measurements}

   \author{Moritz M. M. Pleintinger
          \inst{1}
          \and
          Thomas Siegert\inst{2}
          \and
          Roland Diehl\inst{1}
          \and
          Yusuke Fujimoto\inst{3}
          \and
          Jochen Greiner\inst{1}
          \and
          Martin G. H. Krause\inst{4}
          \and
          Mark R. Krumholz\inst{3,5}
          }

   \institute{Max-Planck-Institut f\"ur extraterrestrische Physik, Gießenbachstraße, 85748 Garching, Germany\\
   \email{mpleinti@mpe.mpg.de}
   \and
	Center for Astrophysics and Space Sciences, University of California, San Diego, 9500 Gilman Dr, La Jolla, CA 92093-0424, USA
   \and
   Research School of Astronomy and Astrophysics, Australian National University, Canberra, ACT 2611, Australia
   \and
   Centre for Astrophysics Research, School of Physics, Astronomy and Mathematics, University of Hertfordshire, College Lane, Hatfield, Hertfordshire AL10 9AB, UK
   \and
   ARC Centre of Excellence for Astronomy in 3D Dimensions (ASTRO-3D), Canberra, ACT Australia}

   \date{Received Month DD, YYYY; accepted Month DD, YYYY}

 
  \abstract
   {The diffuse gamma-ray emission of $^{26}{\rm Al}$ at 1.8\,MeV reflects ongoing nucleosynthesis in the Milky Way, and traces massive-star feedback in the interstellar medium due to its 1\,Myr radioactive lifetime. Interstellar-medium morphology and dynamics are investigated in astrophysics through 3D hydrodynamic simulations in fine detail, as only few suitable astronomical probes are available.
   }
   {We compare a galactic-scale hydrodynamic simulation of the Galaxy's interstellar medium, including feedback and nucleosynthesis, with gamma-ray data on $^{26}{\rm Al}$ emission in the Milky Way extracting constraints that are only weakly dependent on the particular realisation of the simulation or Galaxy structure.}
   {Due to constraints and biases in both the simulations and the gamma-ray observations, such comparisons are not straightforward.
   For a direct comparison, we perform maximum likelihood fits of simulated sky maps as well as observation-based maximum entropy maps to measurements with INTEGRAL/SPI.
   	To study general morphological properties, we compare the scale heights of $^{26}{\rm Al}$ emission produced by the simulation to INTEGRAL/SPI measurements.}
   {The direct comparison shows that the simulation describes the observed inner Galaxy well, but differs significantly from the observed full-sky emission morphology. Comparing the scale height distribution, we see similarities for small scale height features and a mismatch at larger scale heights. We attribute this to the prominent foreground emission sites
   	that are not captured by the simulation.
   }
   {}


   \maketitle
%

\section{Introduction}
	
   $^{26}$Al is an ideal tracer of ongoing nucleosynthesis in the Galaxy. It is produced in massive stars and ejected to their surroundings via stellar winds and \acp{SN}. It decays with a half life time of $\sim$\,0.7\,Myr to $^{26}$Mg and emits a photon at 1809\,keV, which can be measured by gamma-ray telescopes. The spatial distribution of $^{26}$Al provides information about active sites of nucleosynthesis and galactic chemical enrichment, as well as dynamics and feedback processes in the \ac{ISM} throughout the Milky Way \citep[e.\,g.][]{Diehl:2006aa, Wang:2009cj, Diehl:2010aa, Kretschmer:2013aa, Bouchet:2015aa, Siegert:2017aa}.\\   
   Hydrodynamic simulations are a crucial tool for understanding the dynamics and chemical enrichment of galaxies, and for interpreting observations. Empirical models that may describe diffuse radioactivity in the interstellar medium of the Galaxy have a long history \citep[cf.][]{Prantzos:1993mn, Prantzos:1995gd, Knodlseder:1996wh, Lentz:1999ui, Sturner:2001yu, Drimmel:2002il, Alexis:2014ux}. Yet, it is only recent that \citet{Fujimoto:2018aa} reported the first galactic-scale hydrodynamic simulation that starts from basic physical processes and aims to track the synthesis and transport of radioactive isotopes, such as $^{26}{\rm Al}$ and $^{60}{\rm Fe}$ in a Milky Way-like galaxy.\\
   Previous, heuristic model comparisons identify morphological similarities between $^{26}{\rm Al}$ emission and multi-wavelength tracers or geometric emission models \citep[e.g.][]{Hartmann:1994eo, Prantzos:1995gd, Diehl:1997op, Knodlseder:1999we, Diehl:2004ke, Kretschmer:2013aa}, leaving astrophysical implications to their interpretations. It is important to cross-check simulations that are based on astrophysical assumptions with observations. A fundamentally informative comparison of hydrodynamical simulations to actual measurements is challenging because the Milky Way is one particular realisation of a galaxy, and any given hydrodynamic simulation will, even if intended to be similar, not perfectly match it. It is further complicated by the observational limitations of gamma-ray data due to the necessity of image reconstruction methods, compared to direct imaging.\\
   In this paper we investigate a range of methodological approaches for a generalised comparison of $^{26}$Al full-sky emission maps from the simulation performed by \citet{Fujimoto:2018aa} to gamma-ray data measured with the spectrometer SPI \citep{Vedrenne:2003aa} aboard the INTEGRAL satellite \citep{Winker:2003aa}. In Sect.\,\ref{sec:observations_and_simulations} the data analysis procedure for INTEGRAL/SPI observations as well as the properties of the simulation by \citet{Fujimoto:2018aa} are laid out. Different comparison methods are described in Sect.\,\ref{sec:comparison}, concluded by a discussion of the results in Sect.\,\ref{sec:conclusions}.


\section{Observations and simulations}
\label{sec:observations_and_simulations}

\subsection{Gamma-ray measurements}
\label{subsec:measurements}
	
\begin{figure}
		\centering
		\includegraphics[width=\hsize]{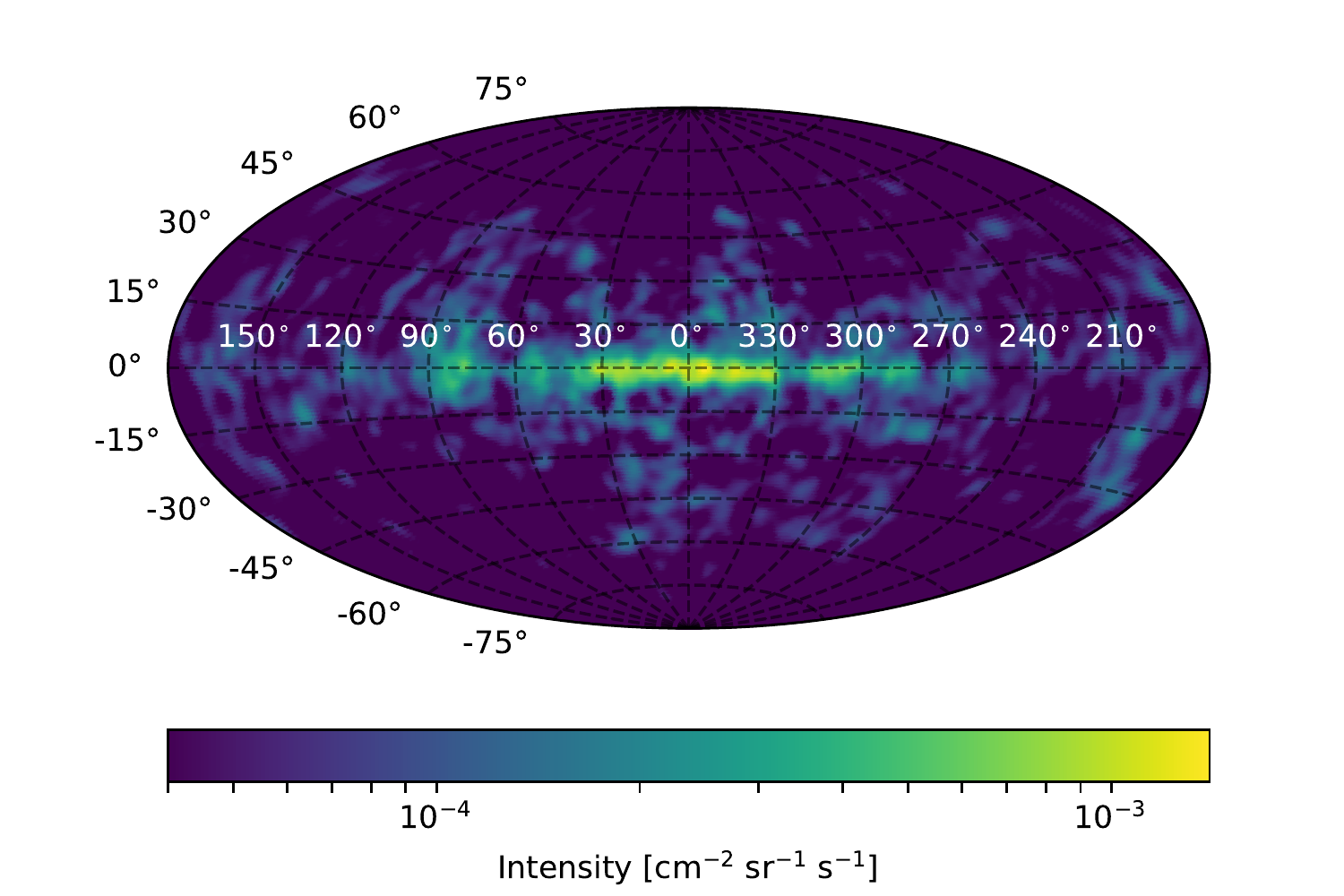}
		\caption{Maximum entropy all-sky map of the 1.8\,MeV emission in the Milky Way obtained from 9\,yr of observations with COMPTEL \citep{Pluschke:2001voa}.}
		\label{fig:comptel_map}
	\end{figure}

\begin{figure}
		\centering
		\includegraphics[width=\hsize]{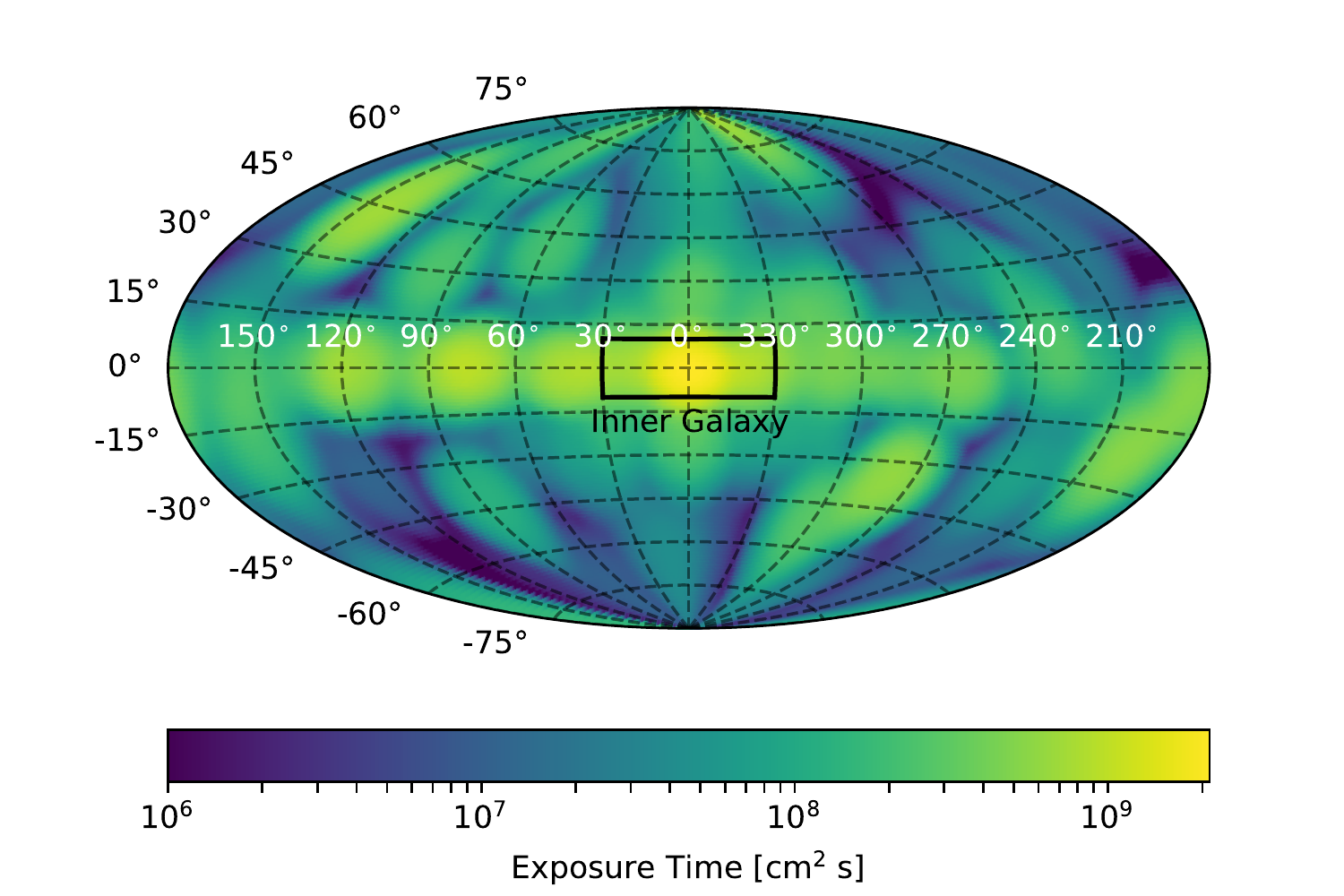}
		\caption{SPI exposure map for 13.5\,yr of the INTEGRAL mission with an integrated observation time of $\sim$\,200\,Ms. The black box encloses the inner Galactic region, which we treated separately in our analysis. The effective area of SPI at 1.8\,MeV is 44\,cm$^2$, accounting for dead detectors \citep{Attie:2003do}.}
		\label{fig:exposure_map}
	\end{figure}

	Imaging with gamma-ray telescopes, either coded-mask based such as SPI or Compton telescopes, such as COMPTEL aboard the CGRO satellite, suffers from the large instrumental background due to cosmic-ray bombardment. Typically, background amounts to 90--99\,\% of measured events in SPI, so that direct imaging is only possible for strong sources. On the other hand, maximum likelihood and maximum entropy approaches are well-tested and can be directly applied to the raw data including an elaborate background model. The first full-sky image of the $1.8$\,MeV emission was obtained by \citet{Oberlack:1996ag} using a maximum entropy deconvolution applied to data from $3.5$\,yr of observations with COMPTEL. In this work, we use the map from the entire nine-year CGRO mission obtained with the same method \citep[Fig.\,\ref{fig:comptel_map},][]{Pluschke:2001voa}. The COMPTEL telescope has an angular resolution of $3.8^{\circ}$. The map shows a large latitude extent on top of a clumpy structure, concentrated to the inner galactic disk region, and associated with spiral arm tangents as well as nearby massive star regions.
	\\
	For our analysis, we use observations of this $^{26}\rm Al$ gamma-ray emission as was subsequently obtained with INTEGRAL, with a SPI data set comprising $\sim$\,$200$\,Ms exposure time from $13.5$\,yr of data in the energy range from $1.795$ to $1.820$\,MeV. We excluded observations with high rates of saturated detector events as well as 20\,\% of the orbital phase around the perigee in order to avoid background from solar flares and passages through the Van Allen radiation belt. The observational sky coverage of the data set in time and space is shown in Fig.\,\ref{fig:exposure_map}.
	The patchy structure comes from the observation strategy of INTEGRAL, observing regions of interest separately, rather than performing a uniform full-sky survey.\\
	In order to spatially resolve this emission, SPI uses a coded-mask technique. It measures gamma-ray energies between 20 and $8000$\,keV with an angular resolution of $2.7$\,$^{\circ}$. According to the morphology of a source or extended emission feature and the orientation of the telescope, a characteristic shadow pattern is projected onto an array of 19 high purity Ge-detectors.
	The celestial signal is overlaid by an instrumental background, originating from nuclear excitations of the instrument and spacecraft material itself. To determine the celestial gamma-ray signal from the shadow pattern, simple background subtraction would lead to erroneous results, as we expect only a small number of counts per energy bin per detector per second and 90\,\% background. Instead, a simultaneous fit for celestial and background signals has to be applied \citep{Strong:2005iy, Diehl:2018aa, Siegert:2019aa}. We use the model description
	\begin{equation}
		m_k = \sum_{j}R_{jk}\sum_{i = 1}^{N_{\text{I}}}\theta_i M_{ij} + \sum_{i = N_{\text{I} + 1}}^{N_{\text{I}} + N_{\text{B}}}\theta_i B_{ik}
		\label{eq:model}
	\end{equation}
	for the measured instrument counts in each particular energy bin $k$ for $N_{\text{I}}$ sky and $N_{\text{B}}$ background model components \citep{Siegert:2019aa}. For each image element $j$, the celestial source model intensities $M_{ij}$ are convolved with the image response function $R_{jk}$, i.e.\ the mask pattern. The parameters $\theta_i$ scale the intensities for all model components $i$. The background contributions $B_{ik}$ are independent of the mask and the spacecraft orientation but vary on different time scales and over the entire mission due to the solar cycle, nuclear build-up processes, solar flares, and radiation belt transits. A detailed background model has been obtained from spectral fitting of the entire mission data \citep{Diehl:2018aa}, which is then adapted to the particular data set subjected to a specific analysis \citep{Siegert:2019aa}. In this adaption, the background component has to be rescaled on an adequate time scale to properly represent such temporal variations. For the $^{26}\rm Al$ analysis, we choose half-year intervals for the background normalisation, in addition to detector failure times \citep[cf.][Sect. 5.1.2]{Siegert:2019aa}. Short-term variations are taken into account and modelled according to the saturated detector events tracking these at high statistical precision \citep[e.g.][]{Jean:2003aa}.\\
	The thickness of the SPI anticoincidene shield, made of $91$ individual BGO crystals, is $5$\,cm. The attenuation cross section for photon energies around 1.8\,MeV in BGO is $4.7 \times 10^{-2}$\,cm$^{2}$\,g$^{-1}$ \citep{Berger:2010iz}. With a density of the crystals of 7.13\,g\,cm$^{-3}$, we estimate that the transmission probability for perpendicularly-incident photons is about 18\%. This leads to an additional signal mainly when the Galactic disk emission is coming from the side. The transmission probability drops below $1$\% for incidence angles larger than $69^{\circ}$; only $32$\% of all pointings are oriented towards these higher latitudes of $|b| > 21^{\circ}$. Thus, the additional signal due to shield transparency can be considered as small. Additionally, this component imprints more signal in outer detectors compared to inner ones. As the orientation of the spacecraft usually remains rather constant during one orbit, this would lead to a characteristic and quasi-constant background detector pattern on that timescale. Such celestial detector count contributions are generally included in our modelled background, because we determine the background and its detector pattern per orbit.\\
	In contrast, the pattern for counts from a sky signal captured within the coded mask varies according to the spacecraft orientation due to the coded mask pattern. This enables us to distinguish background and source contributions.\\
	As the measured detector counts are  Poisson-distributed, the likelihood of a set of model parameters $\bm{\theta}$, given a data set $D$ with $n$ data points is calculated by the full Poisson likelihood
	\begin{equation}
	\mathscr{L}(\bm{\theta}|D) = \prod_{k=1}^{n} \frac{m^{d_k}_k \exp(-m_k)}{d_k!},
	\label{eq:likelihood}
	\end{equation}
	where $d_k$ is the measured number of instrument counts and $m_k$ is the model predicted value as described in Eq. (\ref{eq:model}). To determine the parameter set $\bm{\theta}$ that maximises the likelihood, we use the negative logarithm of the Poisson likelihood dropping the data-dependent term, which is commonly referred to as Cash statistic \citep{Cash:1979aa}
	\begin{equation}
		\log(\mathscr{L}(\bm{\theta}|D)) \approx -2 \sum_{k = 1}^{n}[m_k - d_k \log(m_k)].
		\label{eq:cash}
	\end{equation}
	In our case, full-sky maps are taken as emission models, which are fitted to the data in detector space for each energy bin separately. In order to compare non-nested models, we employ a likelihood-ratio test, which will be described in Sect.\,\ref{subsec:direct_likelihood_comparison}.\\
	We perform a spectral analysis of each fitted model to determine the 1.809\,MeV line flux accurately above Galactic continuum emission which contributes about 5\,\% of the flux in the line band between 1805 and 1813\,keV. We treat the spectral shape as a degraded Gaussian function which includes the effect of charge collection efficiency due to detector worsening over time \citep[e.\,g.][]{Kretschmer:2013aa, Siegert:2017tp, Siegert:2019aa}. The average instrumental resolution at 1.8\,MeV is 3.17\,keV \citep{Diehl:2018aa}.
	
\subsection{Simulated maps}
\label{subsec:maps}

\begin{figure}
		\centering
		\includegraphics[width=\hsize]{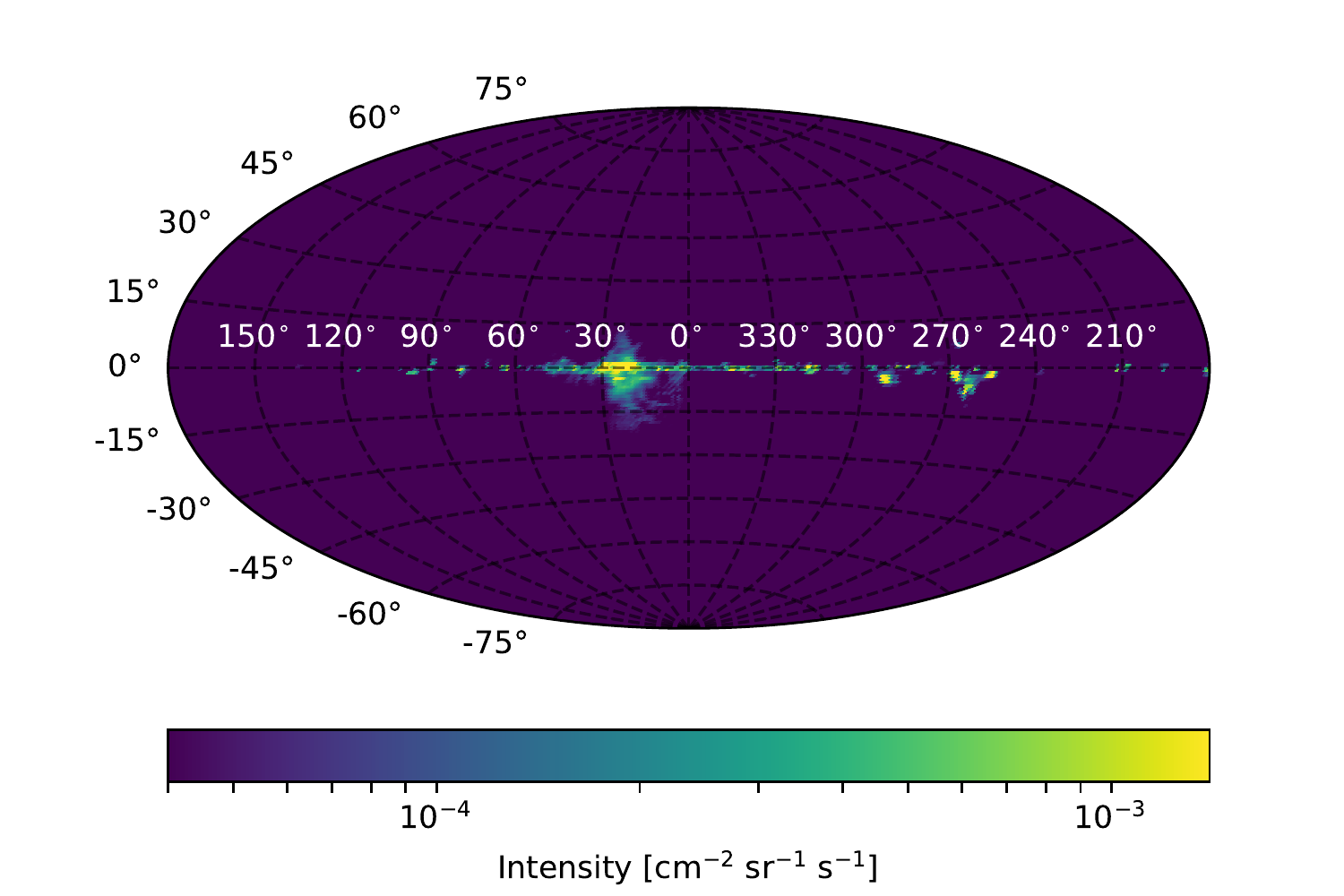}
		\caption{Simulated all-sky flux map of the 1.8\,MeV emission in a Milky Way like galaxy from hydrodynamic simulations by \citet{Fujimoto:2018aa} plotted on the same scale as the COMPTEL map in Fig.\,\ref{fig:comptel_map}.}
		\label{fig:fujimoto_map}
	\end{figure}

	\citet{Fujimoto:2018aa} performed a high resolution hydrodynamic simulation of a Milky Way-like spiral galaxy. They included self-gravity of gas, a fixed axisymmetric logarithmic potential to represent the gravity of old stars and dark matter, radiative cooling, photoelectric heating, as well as stellar feedback in the form of photoionisation, stellar winds, and \acp{SN}. The chemical enrichment of the galaxy was traced by following the dynamics of stellar $^{26}$Al and $^{60}$Fe ejecta, calculated via the \ac{SLUG} population synthesis code \citep{Silva:2012aa, Krumholz:2015aa}. Star-by-star yields of $^{26}{\rm Al}$ are taken from \citet{Sukhbold:2016aa}. For their simulation, they assumed an isolated gas disk orbiting in an otherwise static background potential representing dark matter and a stellar disk component.
	The system evolved for $t = 750$\,Myr with a maximum spatial resolution of $8$\,pc. After that running time, the physical and chemical structure of the ISM had reached a statistical equilibrium characterized by a steady large scale structure of superbubbles filled with the freshly ejected nucleosynthesis products $^{26}$Al and $^{60}$Fe. These bubbles around massive star forming regions were mainly following the spiral arm pattern that developed spontaneously in the simulations, spatially exceeding the size of their host \acp{GMC}. This is consistent with measurements from SPI determining the large scale gas dynamics of $^{26}$Al \citep{Kretschmer:2013aa}.\\
	They derived full-sky flux maps for an hypothetical observer at the Solar Circle at $r = 8$\,kpc by line-of-sight integration of the 1.8\,MeV emission weighted with the distance squared. This was done for 36 different observer positions with 10$^{\circ}$ step size in the plane of the simulated galaxy. These maps cover the whole sky with a pixel size of $1^{\circ} \times 1^{\circ}$. An example of such an integrated map, for an observer at position 0$^{\circ}$, is shown in Fig.\,\ref{fig:fujimoto_map}.
	

\section{Comparison of observations with simulations}
\label{sec:comparison}

\subsection{Direct likelihood comparison}
\label{subsec:direct_likelihood_comparison}
\begin{figure}
		\centering
		\includegraphics[width=\hsize]{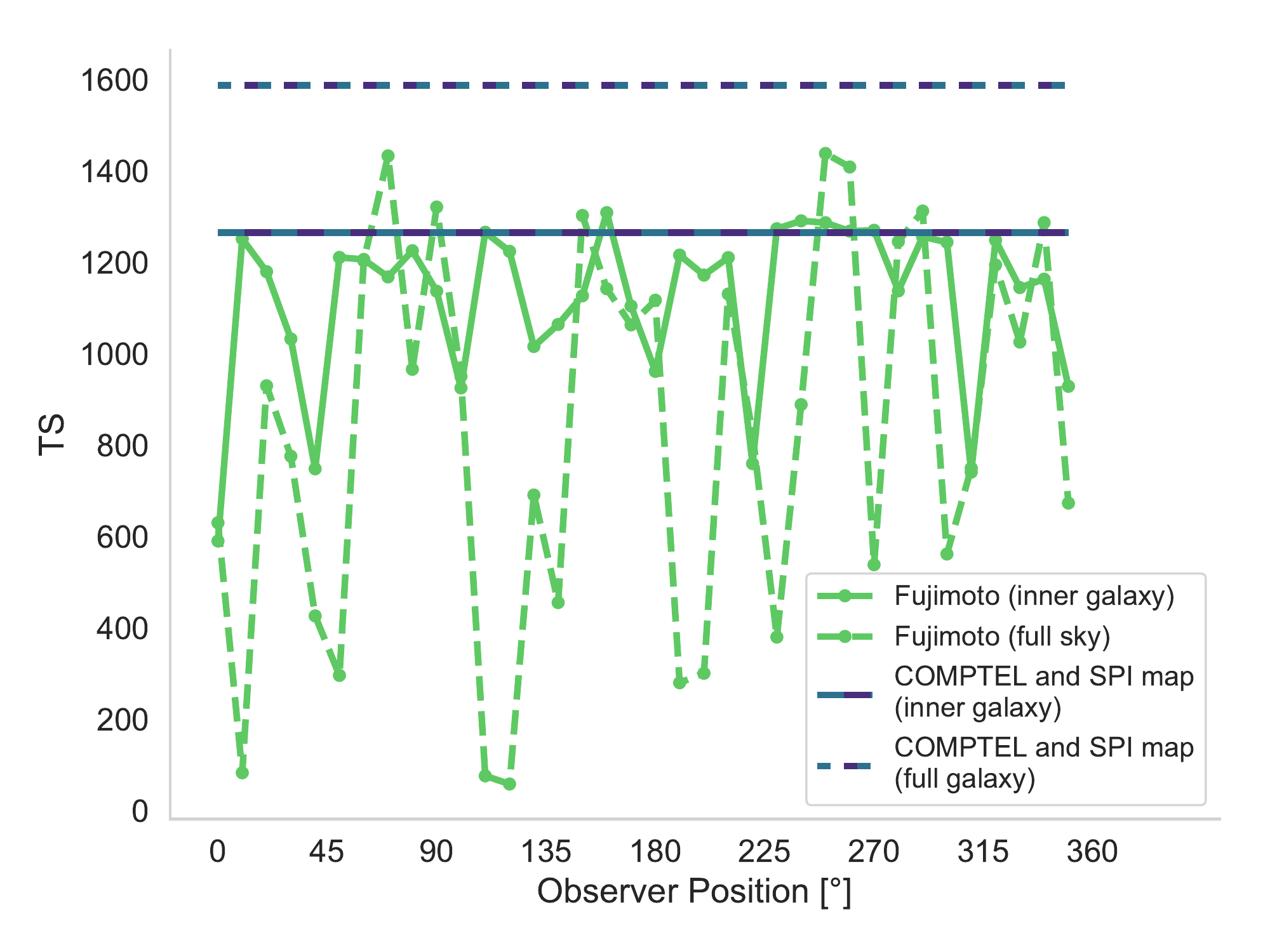}
		\caption{Likelihood ratio of different sky maps relative to the likelihood of a background only fit (see Eq. (\ref{eq:ts})). This is obtained by fitting INTEGRAL data in the energy band 1805\,keV--1813\,keV for the inner Galaxy (solid lines) and the full sky (dashed lines). The simulated sky models (green) correspond to 36 observer positions in the simulation by \citet{Fujimoto:2018aa}. Values of $TS$ for the COMPTEL \citep[][]{Pluschke:2001voa} and SPI \citep[][]{Bouchet:2015aa} maximum entropy maps are given as observation-based reference points. $TS$ values for the COMPTEL and SPI map differ only by 2 (full sky) and 7 (inner galaxy) and overlap in the figure; therefore they are shown combined in single lines (blue/purple). Higher values of $TS$ indicate a lower probability of occurrence by chance in our set of $13.5$ years of SPI data (cf. Appendix \ref{sec:distribution_of_ts}).}
		\label{fig:band}
	\end{figure}

	The most straightforward approach to compare a simulation with observations is a direct evaluation of how well the simulation describes the data. We expect, however, clear deviations due to the specific morphological features of the Milky Way, which are largely random results of the particular distribution of \ac{SN} bubbles around the Sun, which we should not expect a simulation to match in detail. In this approach to comparison, we adopt the 36 flux maps obtained from the simulation by \citet{Fujimoto:2018aa} as emission models for the $^{26}$Al sky. We then perform maximum likelihood fits to the observational datasets using proper instrument response and backgrounds (cf.\ Sect.\,\ref{subsec:measurements}) to determine the likelihood of the fitted simulation map.\\
	Since our statistical method does not provide an absolute goodness of fit, we cannot directly evaluate resulting likelihoods. 
	As a criterion to rate the relative fit quality of different sky models, we apply the test statistic
	\begin{equation}
		TS = \log(\mathscr{L}(M_0|D)) - \log(\mathscr{L}(M_1|D)),
		\label{eq:ts}
	\end{equation}
	which characterises a likelihood-ratio test of a sky model $M_1$ describing celestial emission on top of the background versus the null-model $M_0$ including only the background model. Thus, $TS$ gives the likelihood of $M_1$ relative to the null-hypothesis of observing background only given the data $D$. With a sample of 1000 synthetic Monte Carlo datasets we verify that $TS$ is $\chi^2/2$-distributed and we can associate it with the probability of $M_1$ occurring by chance in $D$ (cf.\ Appendix\,\ref{sec:distribution_of_ts}).\\
	We chose two different realms of this comparison: 
	In the first case, we restrict our analysis to the inner galactic region with $- 10^{\circ} < b < 10^{\circ}$ and $-30^{\circ} < l < 30^{\circ}$ (black box in Fig.\,\ref{fig:exposure_map}), to treat the regions with highest intensity and longest exposure time separately. For the second case, we analyse the full sky emission. For each of the spatial realms, 
	we determine the $^{26}$Al signal in one energy band from 1805\,keV to 1813\,keV.\\
	The model comparison is shown in Fig.\,\ref{fig:band}. As observation-based reference points we include model comparisons with the full-sky maximum entropy map obtained from COMPTEL \citep{Pluschke:2001voa} and a map obtained from SPI measurements \citep{Bouchet:2015aa} as sky models. This gives test statistic values for maps representing observations of the actual Milky Way emission.\\
	For the inner Galaxy, the simulated maps show values of $TS$ mostly below the observation-based maps which are consistent with each other. Nevertheless, there are maps from sight-lines in the simulation which are as unlikely to occur by chance in the data as the observation-based maximum entropy reconstructions.
	This indicates that the observed structure in this region is by and large dominated by the overall Galactic morphology and therefore well described by the generic hydrodynamic simulation. There are surprising variations among the different simulation samples, however. 
	It is particularly striking that there are some observer positions where the simulations are actually less likely to be found by chance in the SPI data than the maximum entropy reconstruction.
	It is difficult to identify distinct morphological features, to which the striking improvement for some of the observer positions may be attributed. There is indication that this may be mainly due to the geometric configuration of superbubbles in the direction of the Galactic centre. We return to this issue in Sect.\,\ref{subsubsec:galaxy_wide_scale_height_and_scale_radius}.\\
	Taking the full sky into account, the observation-based maps show overall larger and also consistent values of the test statistic $TS$, while values for the simulated maps fall below, with a larger scatter than for the inner galaxy case. This indicates less-matching models (through a higher probability of chance coincidence for the simulated maps, compared to the observation-based maps).
	Thus the simulations diverge significantly from the maximum entropy reconstruction over the full sky, particularly when we include higher latitudes.
	\\
	Because the simulation primarily gives relative intensity variations on the sky, we fit the line flux to estimate the absolute changes. The obtained fluxes are given in Table\,\ref{table:fit} in comparison with the observational maximum entropy maps. 
	\begin{table}
		\caption[]{Flux of the 1.8\,MeV line in units of $10^{-3}$\,ph\,cm$^{-2}$\,s$^{-1}$, derived for different morphological maps and spatial scopes. The uncertainties for the simulation denote the 68th percentile of all 36 flux maps from \citet{Fujimoto:2018aa}.}
		\label{table:fit}
		$$
		\begin{array}{p{0.5\linewidth}ll}
		\hline
		\noalign{\smallskip}
		Map      &  \text{Inner Galaxy} & \text{Full Sky}\\
		\noalign{\smallskip}
		\hline
		\noalign{\smallskip}
		COMPTEL  & 0.29 \pm 0.01 				  & 1.71 \pm 0.06 \\
		SPI             & 0.30 \pm 0.01 			     & 2.09 \pm 0.08\\
		Simulation  & 0.26^{+0.03}_{-0.05}       & 0.52^{+0.15}_{-0.21}\\
		\noalign{\smallskip}
		\hline
		\end{array}
		$$ 
	\end{table}
	For the isolated treatment of the inner Galaxy, the line fluxes derived from all maps match within the uncertainties. The flux values obtained from the full-sky analysis are significantly ($\sim$\,7$\sigma$) lower for the simulated maps than for the COMPTEL map. Conversely, the SPI map gives a larger flux ($\sim$\,4$\sigma$). As \citet{Fujimoto:2018aa} state, prominent 1.8\,MeV emission regions such as Cygnus, Carina, or Sco-Cen, as well as the characteristic spiral arm structure of the Miky Way have been omitted in their simulation. Thus, it is reasonable that especially the characteristic foreground emission in the Milky Way, which originates from regions relatively close to the Earth and extends to higher latitudes, is missing. The COMPTEL map as well as the SPI map include such regions and features, thus providing a better fit to INTEGRAL/SPI data. Nevertheless, the good fitting results for the inner Galaxy indicate that this region is less influenced by characteristic foreground features but by the galaxy-wide emission.\\

\subsection{Scale height analysis}
\label{subsec:scale_height}

	Given our finding that numerical simulations fail to mimic the particular structure of the sky as seen from Earth, we now seek a more general method of comparing simulations and observations. Thus, we analyse their morphological features in a generalized way which can be applied to SPI measurements as well. Common morphological analyses like expansion in spherical harmonics or wavelets are very sensitive to the assumptions made for image reconstructions, for example starting points of maximum entropy deconvolution. Thus, we stick to maximum likelihood estimations of chosen models, and evaluate characteristic distributions that can be inferred from those models. A representation of the sky in spherical harmonics has basically an infinite number of realisations for each combination of degree and mode already by simple rotation on the sky. As we can only test each realisation of such an analytic model individually, this is not a basis for a viable approach. Thus, we have to choose a simpler model which contains basic morphological information and for which we can fit individual realisations to the data.

\subsubsection{Galaxy-wide scale height and scale radius}
\label{subsubsec:galaxy_wide_scale_height_and_scale_radius}

\begin{figure}
		\centering
		\includegraphics[width=\hsize]{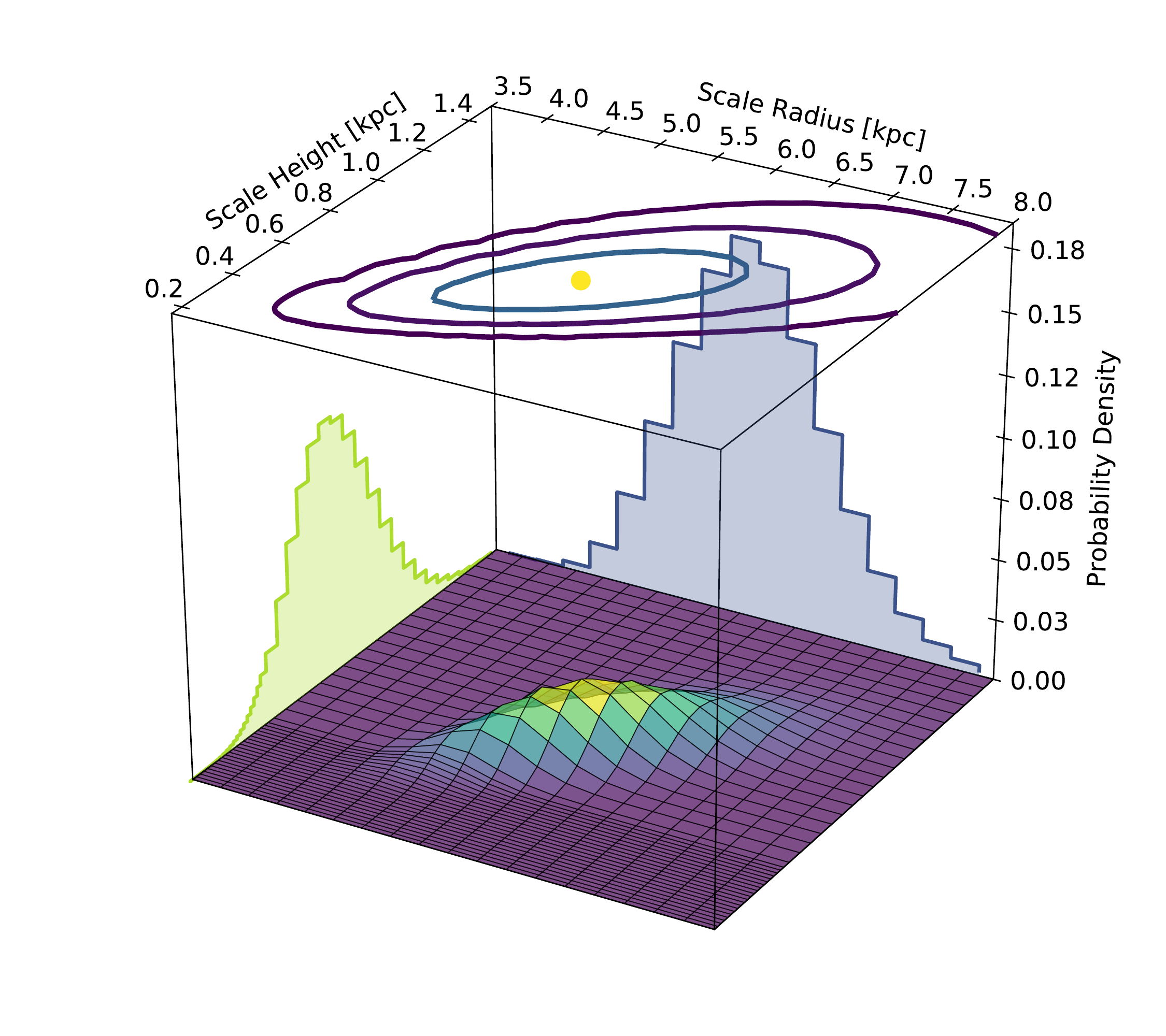}
		\caption{Fit results for $^{26}$Al data described by an exponential disk model. The colour-coded surface gives the \ac{PDF} of scale height and scale radius. The contour plot shows the 1, 2, and 3$\sigma$ uncertainty contours around the best fit values (yellow dot) and the shaded histograms give the likelihood profile for scale height (green) and scale radius (blue).}
		\label{fig:grid}
	\end{figure}

	Extragalactic studies show an exponential decrease of young massive stars with radius. As the 1.8\,MeV emission from $^{26}$Al traces such massive star groups, we expect it to follow a similar trend. Therefore, we assume a doubly exponential disk model
	\begin{equation}
		\rho(r,z) = A_0 \exp\left[ -\left( \frac{r}{r_0} + \frac{|z|}{z_0}\right)\right],
		\label{eq:exponential_disk}
	\end{equation}
	with the galactocentric radius $r^2 = x^2 + y^2$, the height above the disk $z$, scale radius $r_0$, scale height $z_0$, and amplitude of the disk $A_0$, as a first-order model for the galactic 3D distribution of the $^{26}$Al emissivity in units of ph\,cm$^{-3}$\,s$^{-1}$. The 2D flux projection of this emission model onto the celestial sphere in galactic longitude $l$ and latitude $b$ is obtained by line-of-sight integration
	\begin{equation}
		F(l, b) = \frac{1}{4\pi} \int_{s_{\text{min}}}^{s_{\text{max}}} \rho(x_{\text{s}} + s \cdot u_x, y_{\text{s}} + s \cdot u_y, z_{\text{s}} + s \cdot u_z)\ \text{d}s.
		\label{eq:los}
	\end{equation}
	from the relative position of the Sun with respect to the Galactic centre $\vec{p}_{\text{s}} = (x_{\text{s}}, y_{\text{s}}, z_{\text{s}}) = (8.5, 0, 0)$\,kpc along the line-of-sight vector $\vec{u} = (u_x, u_y, u_z) = (\cos(l)\cos(b), \sin(l)\cos(b), \sin(b))$. The integration boundaries $s_{\text{min}}$ and $s_{\text{max}}$ confine the emission model to the volume of the Milky Way. Earlier studies of the inner Galaxy with SPI find a $^{26}\rm Al$ scale height between $60$ and $250$\,pc \citep{Diehl:2006aa, Wang:2009cj} for a fixed scale radius of 4\,kpc. COMPTEL all-sky datasets also suggested a scale height of $150$--$170$\,pc \citep{Oberlack:1997th, Diehl:1997op}. In order to obtain a galaxy-wide evaluation of the $^{26}\rm Al$ emission with SPI, we fit a grid of $32 \times 64$ combinations of different $r_0$ and $z_0$ to our set of $13.5$\,yr of SPI data. The scale radius $r_0$ ranges from 0.50 to 8.25\,kpc in 250\,pc steps. The scale height $z_0$ ranges from 10 to 475\,pc in 15\,pc steps and from 500 to 2050\,pc in 50\,pc steps \citep{Siegert:2017tp}. From the resulting Poisson likelihood values we calculate the probability density distribution for combinations of scale height and scale radius of the galaxy-wide $^{26}$Al emission. The results are shown in Fig.\,\ref{fig:grid}.
	From the likelihood profiles of both parameters, we find a best fitting exponential disk model with $r_0 = 5.81 \pm 0.64$\,kpc and $z_0 = 0.77 \pm 0.17$\,kpc for the Milky Way emission. The smaller scale height of $180$\,pc found with SPI by \citet{Wang:2009cj} is due to a spatial restriction to the inner Galaxy.\\
	We fit the same grid of exponential disk emission models to the 36 simulated $^{26}$Al flux maps. This enables a comparison of the galaxy-wide morphology of simulations and gamma-ray measurements. We evaluate the best fitting amplitude for each exponential disk model and calculate the difference between each simulated map and all exponential disk maps. In order to capture the same dynamical range as in SPI observations, we multiply the simulated sky maps with the SPI exposure map. This transfers the sky maps into count space projected onto the sky with effective area and sky coverage of SPI taken into account. We perform a Pearson's $\chi^2$ fit to estimate the best fitting amplitudes of all exponential disk models for each simulated map, treating the maps from the simulation as synthetic data and the exponential disk maps as model prediction. From the minimum $\chi^2$ values we find the mean values of $r_0 = 3.02$\,kpc
	and $z_0 = 0.07$\,kpc
	for all synthetic sky maps.
	While the overall scale radius in the simulation is close to what we observe in the Milky Way, the scale height appears to be one order of magnitude smaller. However, we find one outlier at an observer position of 100$^{\circ}$ in the simulated galaxy, which shows the maximum overall scale height of $z_0 = 0.7$\,kpc in agreement with observations. In the simulation, this is a unique spot where the observer is placed directly inside a superbubble of $\sim$\,1\,kpc in size around two high star formation clumps, located in the direction of the galactic centre. This could indicate that, from a nucleosynthesis point of view, the Sun inside the Local Bubble is located at such a rather exceptional location in the Galaxy.

\subsubsection{Scale height frequency spectrum}

\begin{figure}
		\centering
		\includegraphics[width=\hsize]{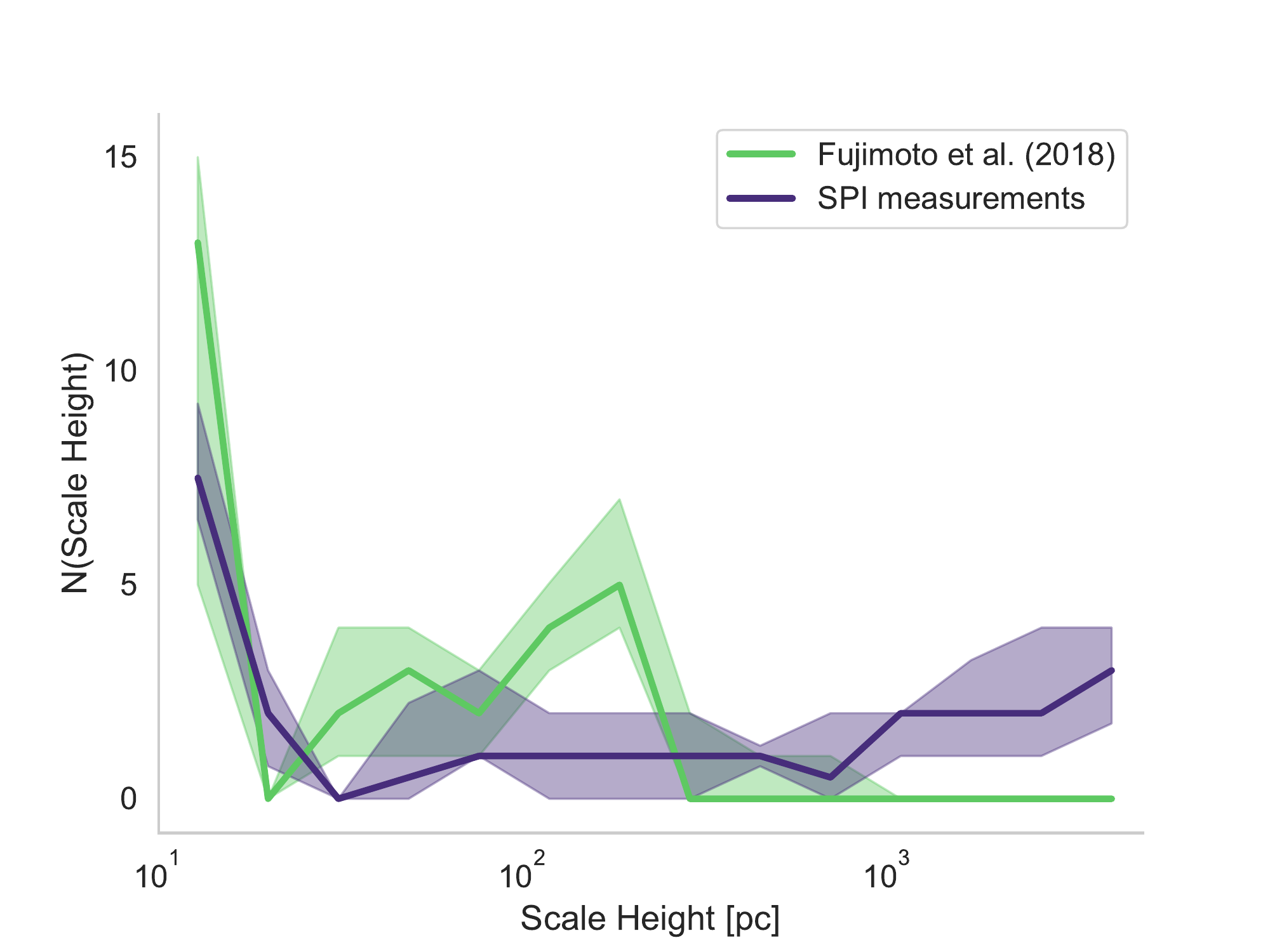}
		\caption{Frequency spectrum of best fitting scale heights in 12$^{\circ}$ longitude bins. The green line shows the median of all 36 flux maps from the simulation by \citet{Fujimoto:2018aa} of which 68\,\% lie inside the shaded green region. The purple line and shaded region give the median and the 68th percentile of SPI observations for 12 sets of longitude bins shifted by 1$^{\circ}$ each.}
		\label{fig:scale_height_spectrum}
	\end{figure}

	In order to spatially resolve how the overall scale height is composed of certain features with different latitude extent, we investigate separate rectangular \acp{ROI} of 12$^{\circ}$ in longitude and 180$^{\circ}$ in latitude each. The width in longitude was chosen to achieve a compromise between spatial resolution and intensity of the $^{26}$Al signal per bin \citep{Kretschmer:2013aa}. We also apply synthetic noise reflecting the dynamical range of the observations by a weighting with the SPI exposure as it was done in Sect.\,\ref{subsubsec:galaxy_wide_scale_height_and_scale_radius}. For simplicity, we assume a fixed scale radius of $r_0 = 5.5$\,kpc for the exponential disk models and retain only $z_0$ as free parameter. We determine the fit quality for a set of 123 full-sky exponential disk models with different scale heights between 10\,pc and 5\,kpc for each \ac{ROI} separately. The step size is 15\,pc from 10 to 475\,pc and 50\,pc from 0.5 to 5\,kpc. The scale height of the best fitting model in each longitude bin gives a measure of the extent of $^{26}$Al emission in latitude. Since SPI has a field of view of $16 \times 16$\,deg$^2$, two adjacent ROIs are overlapping. To account for the variability arising from the arbitrary placement of bins and their overlap, the evaluation was done for 12 different sets of bins shifted in $1^{\circ}$ longitude steps.\\
	This allows for a generalized comparison of the variances present in synthetic and observed maps of the galactic $^{26}$Al emission that is particularly independent of the specific morphology seen in the Milky Way. The scale height frequencies derived from the simulation and SPI data are shown in Fig.\,\ref{fig:scale_height_spectrum}.
	From SPI measurements we see a major contribution in small scale height features of the order of $z_0 \sim 10$\,pc. Additionally, a lower and fairly constant contribution of larger latitude extents is measured with a slight increase even up to a few kpc. This indicates the presence of two quite distinct emission features rather than a uniform exponential morphology with a consistent scale height. This implies that the actual Milky Way emission deviates substantially from an exponential disk model.\\
	The simulation shows a similar excess for small scale heights and a second component of scale heights around $z_0 \sim 100$\,pc. Small scale heights indicate dominant galactic disk emission and large scale heights correspond to dominant nearby regions extending towards higher latitudes or even local emission around the Sun. While the frequency of small scale heights in the simulation matches well with observations, the nearby large scale height component seems to be under-represented. This finding is consistent with the results from the direct comparison in Sect.\,\ref{subsec:direct_likelihood_comparison}.

\subsubsection{Scale height vs. longitude}

\begin{figure}
		\centering
		\includegraphics[width=\hsize]{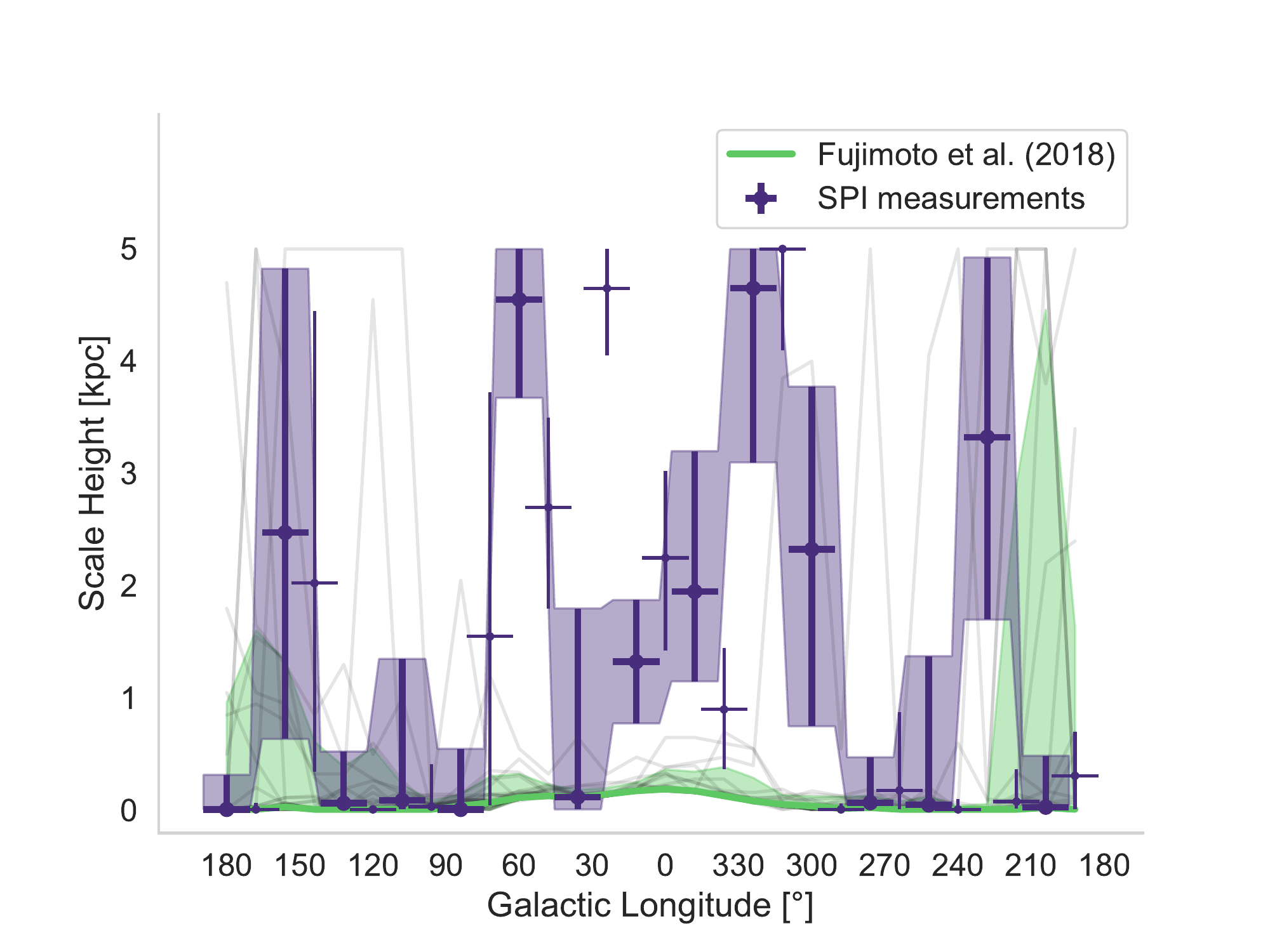}
		\caption{Best fitting scale height for individual 12$^{\circ}$ longitude bins along the galactic plane. The grey lines show the results for all 36 observer positions in the simulation by \citet{Fujimoto:2018aa} individually. Their median is given by the green line with the 68th percentile as shaded green region. The purple crosses represent the maximum likelihood scale heights obtained from SPI data with errors in y-direction depicting the average width of the $\log(\mathscr{L})$-profile in each longitude bin. Since adjacent ROIs are overlapping, data points are shown alternately with bold and thin markers to group mostly-independent data points together. The shaded region follows only the bold data points to guide the eye for a possible trend. The fits are restricted to 5\,kpc scale heights.}
		\label{fig:scale_height_vs_longidtude}
	\end{figure}

	To see how the different scale height components are linked to the galactic morphology, we break down the results according to galactic longitude. This is depicted in Fig.\,\ref{fig:scale_height_vs_longidtude}.
	The uncertainties in scale height represent the average width of the $\log(\mathscr{L})$-profile in each longitude bin. Due to overlapping ROIs neighbouring data points are not strictly independent and grouped together alternately. The error bars in longitude account for the overlap.\\
	In regions looking away from the Galactic centre at $-60^{\circ} > l > 60^{\circ}$, SPI observations are dominated by emission with small scale heights of $\sim$\,10\,pc with a few contributions at intermediate large scale heights. On average, this general structure is also seen in the simulation, indicating that these regions in the Milky Way are shaped according to the stochastic galaxy-wide star formation processes. On the other hand, around the Galactic centre at $-60^{\circ} < l < 60^{\circ}$, the measurements show intermediate to very large scale heights of the order of kpc. The flux maps from the simulation show overall smaller scale heights between $10$\,pc and $100$\,pc in this region. In contrast to the previously considered outer regions, this implies a strong contribution of characteristic nearby foreground emission in the direction around $l = 0^{\circ}$, e.g.\ from the Sco-Cen region at  $l = 350^{\circ}$ \citep{Krause:2018aa}. The structural difference could indicate that the run of star formation rate density with galactocentric radius differs in the simulation from the true conditions in the Milky Way, i.e.\ there may be more star formation locally than the simulation assumes. Another possible reason could be that the clustering of star formation in the simulation could differ from the one in the observation. The clustering of superbubbles affects the vertical flow of nucleosynthesis material, with bigger superbubbles causing stronger vertical flows \citep{Glasow:2013aa}. Additionally, the spiral arms in the Milky Way could be more classical density waves than the spontaneously formed structures in the simulation, perhaps related to external perturbations or the bar. This may concentrate star formation in fewer, more prominent spiral arms in reality.


\section{Conclusions}
\label{sec:conclusions}

	In this paper we present methodological routes for comparison between observed maps of $^{26}{\rm Al}$ decay in the gamma-ray sky and hydrodynamic simulations of a galactic $^{26}{\rm Al}$ distribution. A direct comparison of a generic simulation with observations necessarily yields relatively poor fits because the simulation does not match particular prominent foreground features of the sky as seen from Earth. Nevertheless, we find that the portion of the sky around the Milky Way centre seems to be dominated by the overall galactic emission rather than characteristic foreground structures, and that the simulation provides a reasonably good match to the observed $\gamma$-ray sky in this region. This provides information about the 3D distribution of the $^{26}$Al emission.\\
	For a morphologically more generalised approach, we investigated the latitude extent of the 1.8\,MeV emission as parametrized by the exponential scale height. The rather broad emission with $z_0 = 0.77 \pm 0.17$\,kpc from SPI observations is only seen in one configuration in the simulation. In this case, the observer is placed inside a large superbubble structure filled with freshly produced $^{26}$Al. This may imply that the Sun is placed in a similarly exceptional location in the Milky Way.\\
	In order to spatially resolve certain emission features, we evaluated the scale height extent in 12$^{\circ}$ longitude bins. Analysis of the SPI maps reveals an almost bi-modal distribution, with most longitudes showing a small scale height $\lesssim 50$ pc, but a small number showing extremely large scale heights of a few kpc. These large scale height bins are mainly seen around the direction towards the Galactic centre. The simulation lacks such features on average. Such large latitude extent of $^{26}$Al has to be associated with nearby regions. Thus, we confirm that the $1.8$\,MeV emission contains a significant contribution from nearby superbubbles, e.g.\ from Sco-Cen or regions along the spiral arm tangents \citep[e.g.][]{Rio:1996mv, Diehl:2010aa, Krause:2018aa}. We find indications that one of the most-nearby Sco-Cen $^{26}\rm Al$-filled superbubbles may have overrun the Solar System already and thus contribute an omnidirectional emission component \citep[cf.][]{Krause:2018aa}. This characteristic seen in the observations, and its relative infrequency in simulated sky maps, suggests that the Milky Way may have a more coarse-grained superbubble structure than modelled in the simulation. Indeed, \citet{Fujimoto:2019hg} found that the pre-supernovae stage feedback implemented in \citet{Fujimoto:2018aa} is inefficiently strong to disperse the surrounding gas completely, leaving star formation tracer emission too strongly associated with molecular gas tracer emission, inconsistent with observations of nearby galaxies. Thus, the $1.8$\,MeV map contains important information about the detailed geometry of massive star feedback in the Milky Way.
	The bimodal distribution of scale heights apparent in SPI measurements along different longitudes indicates that Galactic $1.8$\,MeV emission deviates significantly from a simple exponential disk model. This implies that previously determined scale heights to describe the $1.8$\,MeV emission of the entire Galaxy do not correspond to a physical scale height of $^{26}$Al, and instead are probably significantly biased by local foregrounds. Thus, an exponential disk model is insufficient to fit the local scale heights and a purely phenomenological model should be used for this kind of analysis in further studies. This adds systematic uncertainties to the current total $^{26}$Al mass estimate in the Milky Way, because it is based on the assumption of a consistent Galactic scale height \citep{Diehl:2006aa}.\\

\begin{acknowledgements}
	We thank the anonymous referee for the constructive feedback and helpful suggestions on this work. We thank J.\,Michael Burgess for many useful discussions and comments on the statistics part. This research was supported by the Deutsches Zentrum für Luft- und Raumfahrt (DLR). Thomas Siegert is supported by the German Research Society (DFG-Forschungsstipedium SI 2502/1-1). The INTEGRAL/SPI project has been completed under the responsibility and leadership of CNES; we are grateful to ASI, CEA, CNES, DLR, ESA, INTA, NASA and OSTC for support of this ESA space science mission. This research was undertaken with the assistance of resources and services from the National Computational Infrastructure (NCI), which is supported by the Australian Government.
\end{acknowledgements}

\bibliography{bibliography}

\begin{thebibliography}{39}
\expandafter\ifx\csname natexlab\endcsname\relax\def\natexlab#1{#1}\fi

\bibitem[{Alexis {et~al.}(2014)Alexis, Jean, Martin, \&
  Ferri{\`e}re}]{Alexis:2014ux}
Alexis, A., Jean, P., Martin, P., \& Ferri{\`e}re, K. 2014, A{\&}A, 564, A108

\bibitem[{Attie {et~al.}(2003)Attie, Cordier, Gros, Laurent, Schanne, Tauzin,
  von Ballmoos, Bouchet, Jean, Kn{\"o}dlseder, Mandrou, Paul, Roques, Skinner,
  Vedrenne, Georgii, von Kienlin, Lichti, Sch{\"o}nfelder, Strong, Wunderer,
  Shrader, Sturner, Teegarden, Weidenspointner, Kiener, Porquet, Tatischeff,
  Crespin, Joly, Andr{\'e}, Sanchez, \& Leleux}]{Attie:2003do}
Attie, D., Cordier, B., Gros, M., {et~al.} 2003, A{\&}A, 411, L71

\bibitem[{Berger {et~al.}(2010)Berger, Hubbell, Seltzer, Chang, Coursey,
  Sukumar, Zucker, \& Olsen}]{Berger:2010iz}
Berger, M.~J., Hubbell, J.~H., Seltzer, S.~M., {et~al.} 2010, {XCOM: Photon
  Cross Section Database. NIST Standard Reference Database 8 (XGAM)}

\bibitem[{Bouchet {et~al.}(2015)Bouchet, Jourdain, \& Roques}]{Bouchet:2015aa}
Bouchet, L., Jourdain, E., \& Roques, J.-P. 2015, ApJ, 801, 142

\bibitem[{Cash(1979)}]{Cash:1979aa}
Cash, W. 1979, ApJ, 228, 939

\bibitem[{da~Silva {et~al.}(2012)da~Silva, Fumagalli, \&
  Krumholz}]{Silva:2012aa}
da~Silva, R.~L., Fumagalli, M., \& Krumholz, M. 2012, ApJ, 745, 145

\bibitem[{del Rio {et~al.}(1996)del Rio, von Ballmoos, Bennett, Bloemen, Diehl,
  Hermsen, Kn{\"o}dlseder, Oberlack, Ryan, Sch{\"o}nfelder, \&
  Winkler}]{Rio:1996mv}
del Rio, E., von Ballmoos, P., Bennett, K., {et~al.} 1996, A{\&}A, 315, 237

\bibitem[{Diehl {et~al.}(2006)Diehl, Halloin, Kretschmer, Lichti, Schoenfelder,
  Strong, von Kienlin, Wang, Jean, Kn{\"o}dlseder, Roques, Weidenspointner,
  Schanne, Hartmann, Winkler, \& Wunderer}]{Diehl:2006aa}
Diehl, R., Halloin, H., Kretschmer, K., {et~al.} 2006, Nature, 439, 45

\bibitem[{Diehl {et~al.}(2004)Diehl, Kretschmer, Lichti, Sch{\"o}nfelder,
  Strong, von Kienlin, Kn{\"o}dlseder, Jean, Lonjou, Weidenspointner, Roques,
  Vedrenne, Schanne, Mowlavi, Winkler, \& Wunderer}]{Diehl:2004ke}
Diehl, R., Kretschmer, K., Lichti, G., {et~al.} 2004, in Proceedings of the th
  INTEGRAL Workshop on the INTEGRAL Universe, 27--32

\bibitem[{Diehl {et~al.}(2010)Diehl, Lang, Martin, Ohlendorf, Preibisch, Voss,
  Jean, Roques, von Ballmoos, \& Wang}]{Diehl:2010aa}
Diehl, R., Lang, M.~G., Martin, P., {et~al.} 2010, A{\&}A, 522, A51

\bibitem[{Diehl {et~al.}(1997)Diehl, Oberlack, Kn{\"o}dlseder, Bloemen,
  Hermsen, Morris, Ryan, Sch{\"o}nfelder, Strong, von Ballmoos, \&
  Winkler}]{Diehl:1997op}
Diehl, R., Oberlack, U., Kn{\"o}dlseder, J., {et~al.} 1997, in Proceedings of
  the fourth compton symposium (AIP), 1114--1118

\bibitem[{Diehl {et~al.}(2018)Diehl, Siegert, Greiner, Krause, Kretschmer,
  Lang, Strong, Weinberger, Zhang, \& Pleintinger}]{Diehl:2018aa}
Diehl, R., Siegert, T., Greiner, J., {et~al.} 2018, A{\&}A, 661, A12

\bibitem[{Drimmel(2002)}]{Drimmel:2002il}
Drimmel, R. 2002, New Astronomy Reviews, 46, 585

\bibitem[{Fujimoto {et~al.}(2019)Fujimoto, Chevance, Haydon, Krumholz, \&
  Kruijssen}]{Fujimoto:2019hg}
Fujimoto, Y., Chevance, M., Haydon, D.~T., Krumholz, M.~R., \& Kruijssen, J.
  M.~D. 2019, MNRAS, 487, 1717

\bibitem[{Fujimoto {et~al.}(2018)Fujimoto, Krumholz, \&
  Tachibana}]{Fujimoto:2018aa}
Fujimoto, Y., Krumholz, M.~R., \& Tachibana, S. 2018, MNRAS, 480, 4025

\bibitem[{Hartmann(1994)}]{Hartmann:1994eo}
Hartmann, D.~H. 1994, AIP Conf. Proc., 304, 176

\bibitem[{Jean {et~al.}(2003)Jean, Vedrenne, Roques, Sch{\"o}nfelder,
  Teegarden, von Kienlin, Kn{\"o}dlseder, Wunderer, Skinner, Weidenspointner,
  Attie, Boggs, Caraveo, Cordier, Diehl, Gros, Leleux, Lichti, Kalemci, Kiener,
  Lonjou, Mandrou, Paul, Schanne, \& von Ballmoos}]{Jean:2003aa}
Jean, P., Vedrenne, G., Roques, J.-P., {et~al.} 2003, A{\&}A, 411, L107

\bibitem[{Kn{\"o}dlseder {et~al.}(1999)Kn{\"o}dlseder, Bennett, Bloemen, Diehl,
  Hermsen, Oberlack, Ryan, Schoenfelder, \& von Ballmoos}]{Knodlseder:1999we}
Kn{\"o}dlseder, J., Bennett, K., Bloemen, H., {et~al.} 1999, A{\&}A, 344, 68

\bibitem[{Kn{\"o}dlseder {et~al.}(1996)Kn{\"o}dlseder, Prantzos, Bennett,
  Bloemen, Diehl, Hermsen, Oberlack, Ryan, \&
  Sch{\"o}nfelder}]{Knodlseder:1996wh}
Kn{\"o}dlseder, J., Prantzos, N., Bennett, K., {et~al.} 1996, arXiv, 9604053

\bibitem[{Krause {et~al.}(2018)Krause, Burkert, Diehl, Fierlinger, Gaczkowski,
  Kroell, Ngoumou, Roccatagliata, Siegert, \& Preibisch}]{Krause:2018aa}
Krause, M., Burkert, A., Diehl, R., {et~al.} 2018, A{\&}A, 619, A120

\bibitem[{Kretschmer {et~al.}(2013)Kretschmer, Diehl, Krause, Burkert,
  Fierlinger, Gerhard, Greiner, \& Wang}]{Kretschmer:2013aa}
Kretschmer, K., Diehl, R., Krause, M., {et~al.} 2013, A{\&}A, 559, A99

\bibitem[{Krumholz {et~al.}(2015)Krumholz, Fumagalli, da~Silva, Rendahl, \&
  Parra}]{Krumholz:2015aa}
Krumholz, M.~R., Fumagalli, M., da~Silva, R.~L., Rendahl, T., \& Parra, J.
  2015, MNRAS, 452, 1447

\bibitem[{Lentz {et~al.}(1999)Lentz, Branch, \& Baron}]{Lentz:1999ui}
Lentz, E.~J., Branch, D., \& Baron, E. 1999, ApJ, 512, 678

\bibitem[{Mattox {et~al.}(1996)Mattox, Bertsch, Chiang, Dingus, Digel,
  Esposito, Fierro, Hartman, Hunter, Kanbach, Kniffen, Lin, Macomb,
  Mayer-Hasselwander, Michelson, von Montigny, Mukherjee, Nolan, Ramanamurthy,
  Schneid, Sreekumar, Thompson, \& Willis}]{Mattox:1996ai}
Mattox, J.~R., Bertsch, D.~L., Chiang, J., {et~al.} 1996, ApJ, 461, 396

\bibitem[{Oberlack(1997)}]{Oberlack:1997th}
Oberlack, U. 1997, PhD thesis, Technische Universit{\"a}t M{\"u}nchen, Garching
  bei M{\"u}nchen

\bibitem[{Oberlack {et~al.}(1996)Oberlack, Bennett, Bloemen, Diehl, Dupraz,
  Hermsen, Kn{\"o}dlseder, Morris, Sch{\"o}nfelder, Strong, \&
  Winkler}]{Oberlack:1996ag}
Oberlack, U., Bennett, K., Bloemen, H., {et~al.} 1996, A{\&}A Suppl. Ser., 120,
  311

\bibitem[{Pl{\"u}schke {et~al.}(2001)Pl{\"u}schke, Diehl, Sch{\"o}nfelder,
  Bloemen, Hermsen, Bennett, Winkler, McConnell, Ryan, Oberlack, \&
  Kn{\"o}dlseder}]{Pluschke:2001voa}
Pl{\"u}schke, S., Diehl, R., Sch{\"o}nfelder, V., {et~al.} 2001, arXiv, 0104047

\bibitem[{Prantzos(1993)}]{Prantzos:1993mn}
Prantzos, N. 1993, AIP Conf. Proc., 280, 52

\bibitem[{Prantzos \& Diehl(1995)}]{Prantzos:1995gd}
Prantzos, N. \& Diehl, R. 1995, Advances in Space Research, 15, 99

\bibitem[{Siegert(2017)}]{Siegert:2017tp}
Siegert, T. 2017, PhD thesis, Technische Universit{\"a}t M{\"u}nchen, Garching
  bei M{\"u}nchen, Online available at: https://mediatum.ub.tum.de/1340342

\bibitem[{Siegert \& Diehl(2017)}]{Siegert:2017aa}
Siegert, T. \& Diehl, R. 2017, in JPS Conf. Proc. (Journal of the Physical
  Society of Japan), 14--020305

\bibitem[{Siegert {et~al.}(2019)Siegert, Diehl, Weinberger, Pleintinger,
  Greiner, \& Zhang}]{Siegert:2019aa}
Siegert, T., Diehl, R., Weinberger, C., {et~al.} 2019, A{\&}A, 626, A73

\bibitem[{Strong {et~al.}(2005)Strong, Diehl, Halloin, Sch{\"o}nfelder,
  Bouchet, Mandrou, Lebrun, \& Terrier}]{Strong:2005iy}
Strong, A.~W., Diehl, R., Halloin, H., {et~al.} 2005, A{\&}A, 444, 495

\bibitem[{Sturner(2001)}]{Sturner:2001yu}
Sturner, S.~J. 2001, in Proceedings of the Fourth INTEGRAL Workshop, 101--104

\bibitem[{Sukhbold {et~al.}(2016)Sukhbold, Ertl, Woosley, Brown, \&
  Janka}]{Sukhbold:2016aa}
Sukhbold, T., Ertl, T., Woosley, S.~E., Brown, J.~M., \& Janka, H.~T. 2016,
  ApJ, 821, 38

\bibitem[{Vedrenne {et~al.}(2003)Vedrenne, Roques, Sch{\"o}nfelder, Mandrou,
  Lichti, von Kienlin, Cordier, Schanne, Kn{\"o}dlseder, Skinner, Jean,
  Sanchez, Caraveo, Teegarden, von Ballmoos, Bouchet, Paul, Matteson, Boggs,
  Wunderer, Leleux, Weidenspointner, Durouchoux, Diehl, Strong, Cass{\'e},
  Clair, \& Andr{\'e}}]{Vedrenne:2003aa}
Vedrenne, G., Roques, J.-P., Sch{\"o}nfelder, V., {et~al.} 2003, A{\&}A, 411,
  L63

\bibitem[{von Glasow {et~al.}(2013)von Glasow, Krause, Sommer-Larsen, \&
  Burkert}]{Glasow:2013aa}
von Glasow, W., Krause, M., Sommer-Larsen, J., \& Burkert, A. 2013, MNRAS, 434,
  1151

\bibitem[{Wang {et~al.}(2009)Wang, Lang, Diehl, Halloin, Jean, Kn{\"o}dlseder,
  Kretschmer, Martin, Roques, Strong, Winkler, \& Zhang}]{Wang:2009cj}
Wang, W., Lang, M.~G., Diehl, R., {et~al.} 2009, A{\&}A, 496, 713

\bibitem[{Winkler {et~al.}(2003)Winkler, Courvoisier, Di~Cocco, Gehrels,
  Gim{\'e}nez, Grebenev, Hermsen, Mas-Hesse, Lebrun, Lund, Palumbo, Paul,
  Roques, Schnopper, Sch{\"o}nfelder, Sunyaev, Teegarden, Ubertini, Vedrenne,
  \& Dean}]{Winker:2003aa}
Winkler, C., Courvoisier, T. J.~L., Di~Cocco, G., {et~al.} 2003, A{\&}A, 411,
  L1

\end{thebibliography}

\begin{appendix}
\section{Distribution of $TS$}
\label{sec:distribution_of_ts}

\begin{figure}
	\centering
	\includegraphics[width=\hsize]{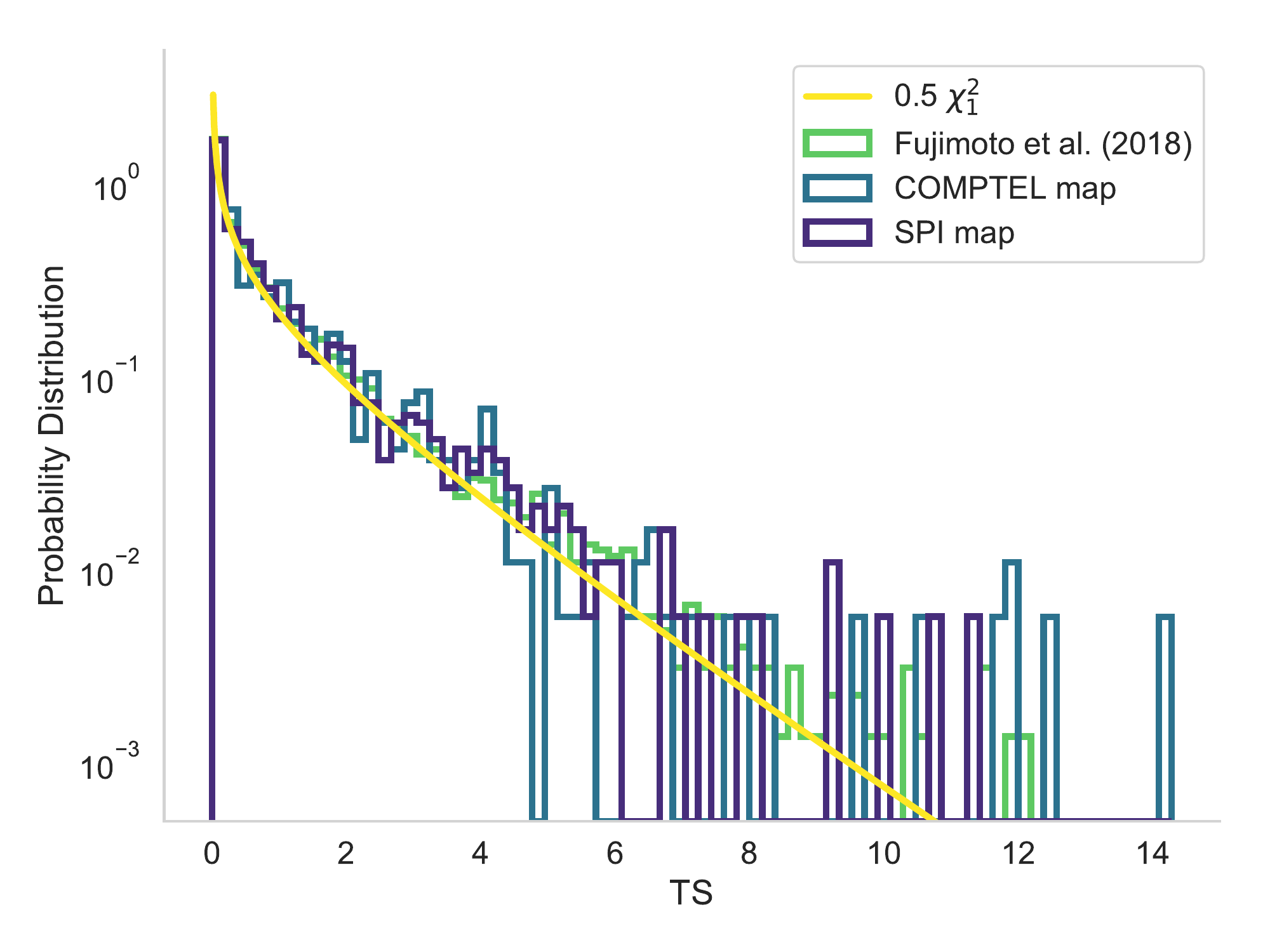}
	\caption{Distribution of the likelihood-ratio test statistic for simulated sky maps by \citet{Fujimoto:2018aa}, the COMPTEL map \citep{Pluschke:2001voa}, and the SPI map \citep{Bouchet:2015aa} according to Eq.\,\ref{eq:ts} using 1000 Poisson datasets sampled from a background model only.}
	\label{fig:test_statistic}
\end{figure}

In order to judge the values of the likelihood ratio $TS$ (cf.\ Eq.\,\ref{eq:ts}) given in Fig.\,\ref{fig:band} with respect to the absolute level of comparison, we investigate its distribution.\\ 
We construct 1000 synthetic Poisson datasets containing background only. Fitting a sky model $M_1$ as well as a background only null-model $M_0$ to those datasets gives $TS$ for occurrence of $M_1$ by chance due to Poisson fluctuations in each dataset. We fitted the COMPTEL map \citep{Pluschke:2001voa}, the SPI map \citep{Bouchet:2015aa}, as well as six cases of outstanding maps by \citet[][observer positions $10^{\circ}$, $70^{\circ}$, $110^{\circ}$, $120^{\circ}$, $250^{\circ}$, and $260^{\circ}$]{Fujimoto:2018aa}. The results are displayed in Fig.\,\ref{fig:test_statistic}, with the \citet{Fujimoto:2018aa} results combined into one average curve. We confirm that $TS$ is $\chi^2_1/2$-distributed for all the maps. The factor 2 is due to the positivity of the Poisson distribution, whereby the negative half of symmetric normal distribution is left out \citep[e.g.][]{Mattox:1996ai}. Thus, we can associate $TS$ with the probability for occurrence by chance of a sky model $M_1$ in our dataset with an upper limit of $10^{-3}$ and evaluate absolute values of the test statistic $TS$.
\end{appendix}

\end{document}